# Sticking coefficient of hydrogen and deuterium on silicates under interstellar conditions

H. Chaabouni[1], H. Bergeron[2], S. Baouche[1], F. Dulieu[1], E. Matar[1], E. Congiu[1], L. Gavilan[1], and J. L. Lemaire[1]

[1] LERMA, UMR 8112 du CNRS, de l'Observatoire de Paris et de l'Université de Cergy Pontoise, 5 mail Gay Lussac, 95031 Cergy Pontoise Cedex, France.
e-mail: Henda.Chaabouni@obspm.fr
[2] Université Paris-Sud, ISMO, UMR 8214 du CNRS, Bat.351, 91405 Orsay, France.
e-mail: herve.bergeron@u-psud.fr

**Abstract**

**Context.** Sticking of H and D atoms on interstellar dust grains is the first step in molecular hydrogen formation, which is a key reaction in the InterStellar Medium (ISM). Isotopic properties of the sticking can have an incidence on the observed HD molecule.
**Aims.** After studying the sticking coefficients of $H_2$ and $D_2$ molecules on amorphous silicate surfaces experimentally and theoretically, we extrapolate the results to the sticking coefficient of atoms and propose a formulae that gives the sticking coefficients of H and D on both silicates and icy dust grains.
**Methods.** In our experiments, we used the King and Wells method for measuring the sticking coefficients of $H_2$ and $D_2$ molecules on a silicate surface held at 10 K. It consists of measuring with a QMS (quadrupole mass spectrometer) the signals of $H_2$ and $D_2$ molecules reflected by the surface during the exposure of the sample to the molecular beam at a temperature ranging from 20 K to 340 K. We tested the efficiency of a physical model, developed previously for sticking on water-ice surfaces. We applied this model to our experimental results for the sticking coefficients of $H_2$ and $D_2$ molecules on a silicate surface and estimated the sticking coefficient of atoms by a single measurement of atomic recombination and propose an extrapolation.
**Results.** Sticking of H, D, HD, $H_2$, and $D_2$ on silicates grains behaves the same as on icy dust grains. The sticking decreases with the gas temperature, and is dependent on the mass of the impactor. The sticking coefficient for both surfaces and impactors can be modeled by an analytical formulae $S(T) = S_0 (1 + \beta T/T_0)/(1 + T/T_0)^\beta$, which describes both the experiments and the thermal distribution expected in an astrophysical context. The parameters $S_0$ and $T_0$ are summarized in a table.
**Conclusions.** Previous estimates for the sticking coefficient of H atoms are close to the new estimation; however, we find that, when isotopic effects are taken into account, the sticking coefficient variations can be as much as a factor of 2 at T = 100 K.

## 1. Introduction

In the InterStellar Medium (ISM), the dust to gas ratio averages 0.01 in mass (Hollenbach & Salpeter 1970). The dust size distribution follows the law proposed by Mathis et al. (1977), which is a power law in nature. In dense clouds, dust is made of grains (covered with icy mantles) having an average size of 0.1 μm (Hollenbach et al. 2009). In diffuse clouds, bare dust grains composed of silicates and/or carbonaceous materials have a size distribution ranging from the smallest 1 nm to 10 nm grains (aromatic hydrocarbon PAHs and amorphous carbons) up to ~ 1 μm sized grains (amorphous silicates) (Mathis et al. 1977; Draine & Lee 1984; Jones 2001). In diffuse regions, grains are known to act as catalysts, helping the production of molecules enriched in hydrogen, such as $H_2$, $H_2O$, $H_2CO$, or $CH_3OH$, therefore the sticking of atoms and molecules is the first step in understanding the chemistry that occurs on the surface of grains. The sticking coefficient S is the probability that a species coming from the gas phase stays on the grain long enough to be bound at a site on the surface.

The sticking coefficient depends primarily on the gas temperature and less on the grain temperature. Nevertheless, the latter is a key parameter of the subsequent chemistry because the density of reactants on the grains depends not only on S, but also on the desorption rate of species (once the species have stuck). If the desorption rate is higher than the accretion rate, the species remain in the gas phase, otherwise they condensate on the grain.

The temperature of the gas affects the velocity of the species (varying as $\sqrt{T}$), and changes the incoming flux by the same factor. In addition, the temperature of the gas also influences the sticking coefficient itself. The sticking process is mostly governed by the ability of a gas species to lose its kinetic energy and become trapped on the surface. Therefore, a high adsorption energy and/or a good momentum transfer will increase the sticking efficiency. The sticking of gas species on a cold surface can be divided into two categories: the light particles (H, $H_2$, D, $D_2$, HD, He) and the heavier molecules or atoms. The light particles, with mass < 4 a.u. in comparison to the mass of atoms and molecules composing the grains (C, O, Si, or $H_2O$), have a small binding energy linked to their weak polarizability. The sticking of heavy elements benefits both from good binding energy momentum transfer and higher adsorption energy, and it is usually assumed that the sticking is close to unity. As proof, experimental studies have shown that the sticking probability of $N_2$, $O_2$, CO, $CH_4$, and $H_2O$ is higher than 90 % (Kimmel et al. 2001; Fuchs et al. 2006; Acharyya et al. 2007).

The sticking of light molecules and atoms is raising numerous questions and has already been the subject of extensive research. A critical problem is the sticking of H atoms, as further hydrogenation of the surface is believed to be a key process for gas grain interaction. The chemical nature of dust grains in the ISM has not been characterized well, but astronomical data show that they are principally composed of silicate and carbonaceous material. In diffuse (low density) clouds, the dust grains are bare, but in dark (dense) clouds the grains are covered with an icy mantle, mainly composed of amorphous solid water with added CO, $CO_2$, methanol, and other molecules (Gibb et al. 2000; Greenberg 2002). Since the formation of $H_2$ molecules (in the ISM) from H atoms is assumed to be preceded by the sticking of hydrogen atoms on dust grains, many theoretical studies devoted to this topic have been developed (Hollenbach & Salpeter 1971; Burke & Hollenbach 1983; Leitch-Devlin & Williams 1985). We cite in particular Buch & Zhang's article (1991), which included the following new elements with respect to the previous studies: (1) detailed modeling of surface structure of amorphous ice, including surface roughness and disorder; (2) a detailed description of intermolecular interactions, including nonsphericity of water molecules. They numerically evaluate the sticking of hydrogen atoms on the cluster of water molecules using molecular dynamical simulations. They proposed the simple formulae: $S = (k_B T/E_0 + 1)^{-2}$, where $E_0/k_B = 102$ K for H atoms and 200 K for D atoms. Al-Halabi et al. (2002) have also calculated the sticking probability of hydrogen atoms to the basal plane (0001) of hexagonal crystalline ice (Ih) surface using classical trajectory calculations and the same H-$H_2$O interacting potential. These theoretical studies show in both cases that the sticking probability of hydrogen atoms decreases with an increasing incident energy $E_i$ of atomic hydrogen. A theoretical study of Cuppen et al. (2010) dedicated to $H_2$ formation on graphitic dust grains in warm conditions complements this work. Modelers have sometimes assumed that the sticking is $\sqrt{(10/T)}$, obtaining therefore a constant formation rate for $H_2$ molecules (Le Petit 2002; Le Petit et al. 2006).

Recently, Matar et al. (2010) have presented an experimental study and a model for the sticking of molecular hydrogen on nonporous amorphous solid water ice held at 10 K. They studied the variation in the sticking coefficient of $D_2$ and $H_2$ molecules as a function of the impinging molecular beam temperature. A theoretical model was developed based upon three assumptions: i) the amorphous structure can be considered as a sum of independent cells, where globally no preferential direction exists; ii) a cell-dependent critical velocity exists below which the atoms or molecules are sticking independently of the interaction details; iii) a probability law specifies the distribution of cell critical velocities. Following this approach, the following analytical formulae (1) gives the thermal sticking coefficient $S(T)$ of a gas at temperature T and provides two physical parameters $S_0$ and $T_0$ to describe the sticking process of the hydrogen species on the grain surface:

$$S(T) = S_0 \frac{(1 + \beta T/T_0)}{(1 + T/T_0)^\beta}. \quad (1)$$

The parameter $S_0$ represents the sticking coefficient of particles at zero temperature ($S_0$ depends on the characteristics of the projectile and the surface in an unspecified way), while the prameter $T_0$ verifies $k_B T_0 = 2mc_0^2$, where m is the mass of the impinging particle, and $c_0$ the mean value of the critical velocities previously mentioned. Under mild assumptions (light impinging particles), $c_0$ not only depends on the different electronic structures of the projectile and the surface but also on the surface temperature. The linear mass dependence of $T_0$ is a critical prediction of the model that can be verified using experimental data obtained with isotopic projectiles. Finally the $\beta$ parameter reflects the geometry of the incident beam. The value $\beta = 2.5$ corresponds to the velocity distribution of a free gas at thermal equilibrium, whereas the value $\beta = 2.22$ corresponds to the velocity distribution of the effusive collimated beam of our experimental setup (Matar et al. 2010).

In this paper, we present the experimental results for the variation in the sticking coefficient of $H_2$ and $D_2$ on a silicate surface held at 10 K as a function of gas temperature, and we discuss the efficiency of the physical model developed by Matar et al. (2010) to fit the experimental data. In section 2, we briefly describe the experimental setup and procedures. In section 3, we present the experimental results and those obtained with the model. In section 4, we analyze and discuss the obtained results, and offer conclusions in section 5.

## 2. Experimental methods

### 2.1. Experimental procedures

The experiments were performed using the FORMOLISM (FORmation of MOLecules in the ISM) setup. It is briefly described here and more details are given in Matar et al. (2010); Lemaire et al. (2010). The apparatus consists of an ultra-high vacuum (UHV) stainless steel chamber with a base pressure lower than $1 \times 10^{-10}$ mbar. In its center, we find the sample holder, which is thermally connected to a cold finger of an Arscryo D210 closed-cycle He cryostat, which can be cooled down to 5.5 K. The temperature is measured in the range of 5.5-350 K with a calibrated silicon diode clamped to the sample and connected to a Lakeshore 336 controller. The temperature is controlled to ±0.2 K with an accuracy of ±1 K. The sample holder is made of a 1 cm diameter copper bloc. Its optically polished and gold-coated surface is covered with an amorphous olivine type silicate film, whose general chemical formula is: $(Fe_x, Mg_{1-x})_2 SiO_4$, where $0 < x < 1$. The silicate sample is provided by Dr D'Hendecourt's group (IAS Orsay) and prepared by the thermal evaporation of San Carlos olivine (Djouadi et al. 2005). Its exact chemical composition is unknown but its amorphous structure, checked by infrared spectroscopy, is characterized by a broad absorption feature around 950 $cm^{-1}$. From the deposition time, the silicate thickness is estimated as ∼ 100 nm fully covering the gold surface.

The surface of our silicate sample is believed to be compact rather than porous, because it shows the same behavior as the surface of nonporous amorphous solid water (np-ASW) ice film grown at 120 K. Such a film has a restricted effective surface area available for adsorption in comparison to a porous amorphous solid water (p-ASW) ice film prepared at 10 K. The surface behavior of our silicate was observed from temperature-programmed desorption (TPD) experiments in our laboratory, which showed similar saturation coverage of these species (CO, $O_2$, $H_2O$, and $D_2$) on both silicate and np-ASW water-ice surfaces held at 10 K using the same amount of flux. On the other hand, according to Amiaud et al. (2007) and Fillion et al. (2009), results for amorphous water ices, the molecular saturation of the compact np-ASW ice surface occurs close to 0.5 ML exposure of $D_2$ at 10 K (1 ML = $10^{15}$ molecules.$cm^{-2}$), corresponding to a few seconds $D_2$ deposition time, whereas p-ASW ice surface with a wide distribution of adsorption sites requires longer exposure times and a higher saturation dose of $D_2$ (about 4 ML). The compact structure of our silicate sample has also been confirmed by the King and Wells experiments (described in the next later section), which indicate a gradual saturation of the silicate surface after ∼ 100 seconds irradiation of $D_2$ at 10 K as in the case of the np-ASW ice film (Matar et al. 2010, Fig 3) for the same amount of $D_2$ flux ~ $9\times10^{12}$ molecules.cm$^{-2}$.s$^{-1}$ (Lemaire et al. 2010).

The hydrogen and deuterium molecules are introduced into the UHV chamber via a triple differentially pumped molecular beam line aimed at the sample holder at an incidence angle of 62°. The gas flows through an aluminum accommodator, connected to an Arscryo D202, closed-cycle He cryostat. The gas is cooled down to a controlled temperature T before entering the UHV chamber. It is therefore possible to vary the gas temperature from 20 K to 350 K. A valve located between the second and the third stages of the beam line (separated by a 3 mm diaphragm) is used to create an effusive beam. At the entrance of the main chamber, also separated from the third stage by a 3 mm aperture, a flag is used either to intercept the beam, in order to estimate the background pressure characterizing the beam, or to allow the species in gas phase to directly reach the surface of the sample holder. An analytical QMS (quadrupole mass spectrometer) Hiden 3F is located above the sample holder and is used to measure the signals of hydrogen species ($H_2$, $D_2$) diffused by the surface during the molecular exposure of the silicate sample.

## 2.2. Sticking coefficient measurements

All our sticking coefficient measurements for $D_2$ and $H_2$ on a silicate surface held at 10 K were performed using the beam reflectivity King & Wells (1972) technique. It consists of measuring with the QMS the background partial pressure of $D_2$ (or $H_2$) molecules in the main chamber during the exposure of the cold surface to the $D_2$ or $H_2$ beam. This experiment is repeated at several beam temperatures ranging from 20 K to 340 K in order to study the variation in the sticking coefficient of hydrogen species on the silicate surface as a function of the temperature of the impinging molecules. Figure 1 shows the $D_2$ normalized signals as a function of the exposure time for different $D_2$ beam temperatures, ranging from 20 K to 340 K.

In these experiments, exposure begins at t = 0 when the $D_2$ beam is directed at the surface in the main chamber. At this time the surface is considered to be free of $D_2$ molecules. All exposure experiments last about 500 s until saturation of the surface by $D_2$ molecules. As shown in Figure 1, during the first 100 s of $D_2$ exposure, we observe a linear decrease in all $D_2$ signals and then a rapid rise that reaches the same plateau at about 300 s. Then after further irradiation, the signals start to rise because of the decrease in the sticking coefficient. This rise is obviously due to molecules that begin to desorb from the surface because of their short residence, which approaches to the time between two arrivals of impinging molecules (Amiaud et al. 2007). The plateau that starts at 300 s corresponds to the steady-state regime where the number of sticking molecules becomes equal to that of desorbing ones.

With a silicate surface at a higher temperature (36 K) and a beam temperature of 50 K, it has been shown (Lemaire et al. 2010, Fig. 3) that a steady state is immediately reached as soon as the beam is aimed at the surface because of the low residence time of the molecules at high surface temperatures.

The apparent sticking coefficient is defined from the normalized curves of Figure 1 as the ratio between the amount of $D_2$ that is stuck on the surface and the total incoming $D_2$ molecules. From Figure 1 we can deduce the absolute sticking coefficient S (T) of $D_2$ molecules at t = 0 when the surface of the silicate is free from $D_2$. As shown in Figure 1, the absolute sticking of $D_2$ decreases when the beam temperature increases from 20 K to 340 K. This behavior has been previously observed on the np-

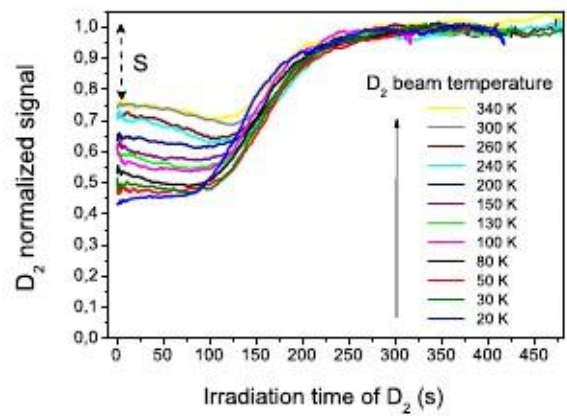

Figure 1. Normalized $D_2$ QMS signals monitored in the vacuum chamber during the irradiation of the silicate sample held at 10 K for several $D_2$ beam temperatures ranging from 20 K to 340 K.

ASW ice by Matar et al. (2010), and it has been explained by the fact that molecules coming from the gas phase at higher beam temperature (and then higher initial kinetic energy) cannot be thermalized efficiently with the cold surface as their excess energy liberated during the collision is not completely transferred to the surface. The thermalization of the newly formed hydrogen molecules to the surface temperature, instead of their prompt desorption into the gas phase, has been discussed by Hornekær et al. (2003) and Congiu et al. (2009) respectively for HD and $D_2$ molecules formed by atomic recombination on porous-ASW ices and by Perets et al. (2007) for $H_2$ formed on amorphous bare silicates.

At almost every beam temperature (except 20 K), the decrease in the signals during the first 100 s of $D_2$ irradiation corresponds to an increase in the sticking coefficient of $D_2$ on the cold silicate surface. This effect is probably induced by the increasing number of $D_2$ adsorbed molecules on the surface that help the thermalization of impinging molecules until some gradual saturation after ~ 100 seconds irradiation (Govers et al.1980). In the case of an impinging gas at very low beam temperature (20 K), the sticking coefficient of $D_2$ molecules is already at its maximum and then remains constant during the first 100 seconds of the irradiation phase.

## 3. Results

The top panel of Figure 2 shows the experimentally found sticking coefficients of $H_2$ and $D_2$ as a function of the molecular beam temperature and the fits obtained from analytical formulae (1). As it shows, the fits to our data are quite satisfactory, especially for the $H_2$ and $D_2$ beam temperatures above 50 K where the sticking coefficient of $D_2$ is higher than for $H_2$. The two physical parameters $S_0$ and $T_0$ for the sticking coefficients, obtained from these fits on silicate surfaces, are $S_0(H_2) = 0.95$, $T_0(H_2) = 56$ K for $H_2$ molecules and $S_0(D_2) = 0.82$; $T_0(D_2) = 112$ K for $D_2$ molecules. The parameters $S_0(H_2)$ and $S_0(D_2)$ satisfy the gas-temperature dependence relation: $T_0(D_2) = 2 \times T_0(H_2)$ as in the case of the np-ASW ice (Matar et al. 2010), where the factor 2 is the mass ratio between $D_2$ and $H_2$ molecules.

To test the validity of the mass dependence of the model (independently of any explicit formulae), we applied the scaling

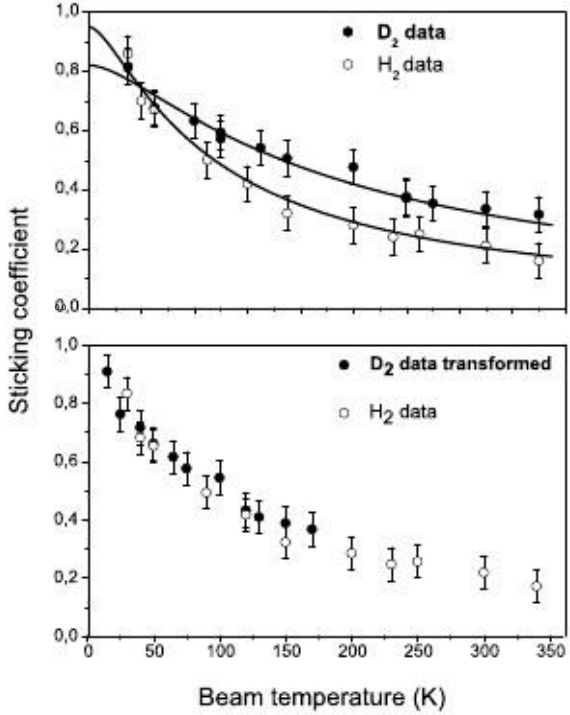

Figure 2. Top panel: variation in the experimental sticking coefficient for $D_2$ (•) and $H_2$ (∘) molecules on the silicate surface held at 10 K as a function of the beam temperature, ranging from 20 K to 340 K. The absolute uncertainties are equal to ±0.06; Fits for $D_2$ and $H_2$ molecules with solid lines, obtained with formulae (1) using $\beta(62°) = 2.22$.
Bottom panel: test of the model using the scaling law of Eq. (2). (∘) $H_2$ data as above, (•) $D_2$ data transformed according to Eq. (2).

law (or renormalization-dilation transformation) given by Eq. (2) (Matar et al. 2010, Eq.12).

$$S_{H_2}(T/2) = \frac{S_0(H_2) \times S_{D_2}(T)}{S_0(D_2)}. \quad (2)$$

The linear mass dependence of the parameter $T_0$ for isotopic molecules implies that the sticking curves for different isotopic species are related by a rescaling on both the S-axis (by the ratio of $S_0$) and on the T-axis (by the mass ratio). Equation (2) specifies this rescaling. Assuming a numerical value of the ratio $S_0(D_2)/S_0(H_2)$ and taking the experimental values for $S_{D_2}(T)$ into account, Eq. (2) allows us to predict new values for the sticking coefficient $S_{H_2}$. The latter can be compared with already known experimental values of $S_{H_2}$: the correlation is presented in the bottom panel of Figure 2 for a fitted ratio $S_0(D_2)/S_0(H_2) = 0.86$. The new data correlates well with the experimental ones especially for the range of gas temperatures $T > 45$ K. We observe that the ratio $S_0(D_2)/S_0(H_2) = 0.86$ is slightly lower than what is obtained for np-ASW ice $S_0(D_2)/S_0(H_2) = 1.10$ (Matar et al. 2010). The value for np-ASW ice is very reliable $S_0(D_2)/S_0(H_2) = 1.10 \pm 0.03$, while the uncertainty is higher in the case of silicate $S_0(D_2)/S_0(H_2) = 0.86 \pm 0.05$. Nevertheless, this is sufficient to say that in the case of np-ASW ice $S_0(D_2)/S_0(H_2) \geq 1$, whereas in the case of silicates $S_0(D_2)/S_0(H_2) \leq 0.91$.

This numerical analysis is accurate enough to prove unambiguously (assuming that the data are not biased) that (a) the ratios $r = S_0(D_2)/S_0(H_2)$ for np-ASW ice and silicates are different, and (b) the isotopic effects $r - 1$ between $H_2$ and $D_2$ have opposite signs for np-ASW ice and silicate surfaces. The structure and the chemical nature of the surface(s) (ice or silicate) are certainly involved in these results, but no simple enlightening argument justifies the strength and the sign of $r - 1$. The sticking probabilities at very low gas temperature involve purely quantum effects that are not easily predicted. We can merely argue that there is no reason to obtain the same value of $r - 1$ when the structure and the chemical nature of the surface are modified.

We can conclude from these results that the transformation law applies to the silicate surface at 10 K and that the isotopic effect between $D_2$ and $H_2$ is explained well by factor 2 corresponding to the mass ratio between $D_2$ and $H_2$ molecules. This confirms that the mass is the most important physical parameter that governs the sticking of light species, regardless of the amorphous surface being studied.

## 4. Analysis and discussion

### 4.1. The molecular case

As observed in both panels of Figure 2, the model reproduces efficiently the experimental data for the sticking coefficient of $D_2$ and $H_2$ on the silicate substrate held at 10 K and especially for beam temperatures higher than 40 K. The experimental sticking coefficients values of $D_2$ and $H_2$ measured on the silicate surface held at 10 K for an impinging gas at T = 300 K are:

$S(D_2) = 0.34 \pm 0.06$ and $S(H_2) = 0.19 \pm 0.06$. These results are slightly lower than those of Matar et al. (2010) on the np-ASW ice for the same molecular beam temperature ($S(D_2) = 0.42 \pm 0.06$ and $S(H_2) = 0.28 \pm 0.06$). The same behavior is, of course, observed for the sticking coefficient values from the model, particularly for $D_2$ molecules, where $S(D_2) = 0.32$ on the silicate and $S(D_2) = 0.43$ on the np-ASW ice (Matar et al. 2010). The mass ratio of the gas phase species and the solid substrate is higher in the case of silicates than in the case of water, therefore the transferred momentum energy is expected to be lower. This is reinforced by the fact that the rigidity (binding energy of the solid network) is also greater in the case of a silicate substrate. As the sticking coefficient reflects the ability of the impactor to lose a fraction of its initial kinetic energy, a lower sticking coefficient in the case of silicates appears to be reasonable. Using this model and related formulae (1), we also extrapolated the sticking parameters $S_0$ and $T_0$ for molecules. The different values of $S_0$ and $T_0$ for molecules are summarized in Table 1. The values for HD are obtained using the temperature relation proportionality, $T_0(HD) = (3/2) \times T_0(H_2)$, which is deduced from the equivalent mass relation proportionality between HD and $H_2$ molecules. The value of $S_0(HD)$ was estimated by an average between $S_0(H_2)$ and $S_0(D_2)$.

### 4.2. The atomic case

The situation for (H and D) atoms is very different from molecules on both theoretical and experimental levels because few calculations are available and atomic detections are difficult.

#### 4.2.1. The experimental situation

Experimental studies for the interaction between H or D with silicate surfaces exist in the literature (Pirronello et al. 1997; Perets et al. 2007; Vidali et al. 2007, 2009), but no experimental estimates for the sticking are indicated. In addition, the assumption of no isotopic effects is usually made.

The sticking of hydrogen atoms is by far harder to solve experimentally, because (a) atoms that are stuck can react and form molecules, this process leads to some difficulties for unbiased measurements within our experimental setup, (b) at the same time it is also known that the already adsorbed molecules play a role in the sticking and the dynamic recombination (Schutte et al. 1976; Govers et al. 1980; Govers 2005; Lemaire et al. 2010). Consequently, the sticking coefficient of hydrogen atoms cannot be directly measured as for molecules ($H_2$ and $D_2$). It can be roughly extrapolated (at least for D) from the recombination efficiency (the probability that two atoms form a molecule).

This extrapolation was derived from our measurements using the same conditions as in Amiaud et al. (2007) on ASW ice, but in the present case, they were done on a silicate substrate held at 10 K and for an atomic beam at 300 K. This procedure leads to the confidence interval S (D) = 0.18 ± 0.08. The lower value S (D) = 0.10 corresponds to the recombination fraction of D atoms at the beginning of the irradiation of the surface with the atomic D-beam, and the upper one S (D) = 0.26 is deduced from the recombination fraction of D atoms when the surface is covered with the nondissociated $D_2$ molecules coming from the atomic beam. The experimental recombination fraction of D atoms on this silicate are not directly measured from our $D_2$ signal, monitored with the QMS during the exposure of the surface to a D atomic beam. However, we should subtract the contribution of the nondissociated part of $D_2$ molecules that are still present in the beam during the discharge (On) and which represents about 35 % of the total molecules (Off). Then, using the signal of $D_2$, resulting from the contribution of the D atomic fraction (65 %), we can estimate the recombination efficiency of D atoms on a silicate surface held at 10 K. This physical magnitude is defined as the fraction of the adsorbed hydrogen atoms coming out as molecules (Govers 2005; Vidali et al. 2007), and it corresponds to the ratio of the total number of D atoms forming $D_2$ molecules during the irradiation phase to the total number of D atoms that impinge on the surface. Thus, the total amount of D atoms exposed on the surface is estimated from the dissociated fraction of $D_2$ molecules present on the surface at the steady state conditions (Perets & Biham 2006; Amiaud et al. 2007; Vidali et al. 2009). Moreover, for low D-atom coverage, the sticking coefficient of D atoms is assimilated into the recombination efficiency if two D atoms coming from the gas phase thermally accommodate on the surface of the silicate at 10 K. They can then diffuse, thanks to their mobility at this surface temperature (Matar et al. 2008), recombine, and form $D_2$ molecules by the Langmuir-Hinshelwood (L-H) mechanism. The increasing adsorbed fraction of the nondissociated $D_2$ molecules until saturation, during the exposure of the silicate surface to the D atomic beam enhances the sticking coefficient of D atoms and increases the recombination efficiency of atoms on the surface (Govers et al. 1980; Amiaud et al. 2007).

### 4.2.2. The theoretical situation

Except for the simulations of Burke & Hollenbach (1983), no theoretical values of the sticking coefficient for H atoms on silicate surfaces are available. Moreover, we suspect an omission to the gas temperature of their coefficient because of the discrepancy between their prediction for the sticking coefficient on water ice (held at 10 K) and the one obtained by Buch & Zhang (1991). For example, assuming a gas temperature T = 100 K, Burke & Hollenbach (1983) predict a sticking coefficient S = 0.7, while Buch & Zhang (1991) predict S = 0.4. As mentioned by Buch & Zhang (1991), the novelty of their

Table 1. Sticking parameters $S_0$ and $T_0$ for $H_2$, $D_2$, H, D, and HD on np-ASW ice (Matar et al. 2010) and on silicate surfaces held at 10 K.

| Substrates | Species | $S_0$ | $T_0$ (K) | References |
|---|---|---|---|---|
| np-ASW ice | H | 1 | 52 | 1 |
| | D | 1 | 104 | 1 |
| | $H_2$ | 0.76 | 87 | 2 |
| | $D_2$ | 0.80 | 174 | 2 |
| | HD | 0.83 | 130.5 | Prediction |
| Silicate | H | 1 | 25 | Extrapolation |
| | D | 1 | 50 | Extrapolation |
| | $H_2$ | 0.95 | 56 | This study |
| | $D_2$ | 0.82 | 112 | This study |
| | HD | 0.87 | 84 | Prediction |

References. (1) Buch & Zhang (1991); (2) Matar et al. (2010).

simulations was to consider the amorphous nature of the surface (including surface roughness and disorder) and a detailed description of intermolecular interactions. No simulation that takes these important physical conditions into account is currently available in the case of silicate surfaces.

### 4.2.3. Modeling the atomic sticking coefficients

The parameters $S_0$ (H) and $S_0$ (D) can be roughly assumed to be unity as in the fits of atomic data by Buch & Zhang (1991) on amorphous water ice (Matar et al. 2010). Moreover, the parameter $T_0$ (H) can be obtained from the mass dependence $T_0$ (D) = 2 × $T_0$ (H). Thus, only the parameter $T_0$ (D) needs to be deduced from experiments to have access to all atomic parameters for H and D.

We therefore assume that the ratio $T_0$ ($D_2$)/$T_0$ (D) ≃ 1.67 obtained in the case of ASW ice (Matar et al. 2010) is roughly unchanged in the case of silicate surfaces. Using our value of $T_0$ ($D_2$), we first deduce $T_0$ (D), then we find S (D) ≃ 0.26 for T = 300 K with our model. This result corresponds precisely to the upper value of S (D) mentioned previously. The rough value $T_0$ ($D_2$)/$T_0$ (D) ≃ 1.67 leads to the required order of magnitude for S (D) at T = 300 K, indicating therefore that the model works well even in the case of an extrapolation to atomic particles. Unfortunately, our assumption about the ratio $T_0$ ($D_2$)/$T_0$ (D) for amorphous water ice and silicates is only a rough approximation. This ratio is expected to be at least weakly dependent on the microphysics, and its value can be different for water ice and silicate substrates. Consequently, the value of $T_0$ (D) cannot be certified.

We thus assume in the sequel that the value of the sticking coefficient of D atoms for an atomic beam at 300 K is

S (D) = 0.18 ± 0.08 (average of our experimental estimates).

We can therefore notice that for an impinging D or $D_2$ gas at room temperature (T = 300 K), S (D) < S ($D_2$) on a cold-silicate surface. This result has been already anticipated on ASW ice substrate and explained by an increase in the elastic scattering of the D atoms relative to $D_2$ (Hornekær et al. 2003).

Assuming $S_0$ (D) = 1 and knowing S (D) from the experiments, we can invert Eq. (1) to obtain $T_0$ (D). We deduce the average value of the parameter $T_0$ for D atoms: $T_0$ (D) = (50 ± 20) K. In this case, the ratio $T_0$ ($D_2$)/$T_0$ (D) on silicates is equal to 2.2 and is slightly higher than that on ASW ice. Once this parameter $T_0$ (D) is known, the parameter $T_0$ (H) is then derived from $T_0$ (H) = (1/2) × $T_0$ (D). The result is $T_0$ (H) = (25 ± 10) K.

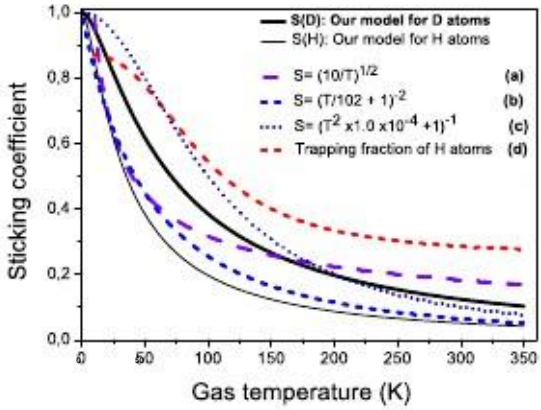

Figure 3. Sticking coefficients on a silicate surface held at 10 K obtained from the model. Thick black line for D atoms and thin black line for H atoms with $T_0(D) = 50$ K and $T_0(H) = 25$ K, respectively, and $\beta = 2.5$ ($\beta$-value for a gas phase at thermal equilibrium). Blue-dashed line obtained from $\sqrt{(10/T)}$ formulae (a) (Le Petit et al. 2006), blue short-dashed line obtained from $(T/102 + 1)^{-2}$ formulae (b) (Buch & Zhang 1991; Andersson & Wannier 1993), blue dotted line obtained from $(T^2 \times 1.0 \times 10^{-4} + 1)^{-1}$ formulae (c) (Le Bourlot 2002) and red dashed line (d) from Burke & Hollenbach (1983)'s theoretical simulations.

Using the previous parameters $S_0(D)$, $T_0(D)$, $S_0(H)$, $T_0(H)$ and assuming the gas phase at thermal equilibrium ($\beta = 2.5$ in formulae (1)), we can plot the sticking curves $S_D(T)$ and $S_H(T)$ for D and H atoms on the silicate surface held at 10 K. This is shown in Figure 3 where three formulaes used in some models to describe the sticking of hydrogen atoms on grain surfaces are also represented: $\sqrt{10/T}$ formulae (Le Petit 2002; Le Petit et al. 2006), $(T/102 + 1)^{-2}$ formulae (Buch & Zhang 1991; Andersson & Wannier 1993), and $(T^2 \times 1.0 \times 10^{-4} + 1)^{-1}$ formulae (Le Bourlot 2002). These results have the expected decrease with some noticeable difference, but the curve $\overline{(T/102 + 1)^{-2}}$ offers the better match. In the case of the $\sqrt{(10/T)}$ formulae, the match with our new estimation is rather good for $T < 50$ K, and for higher gas temperature, the sticking coefficient seems to be overestimated.

In comparison, Figure 3 also shows the trapping fraction curve of H atoms on a silicate surface held at 10 K, estimated from Burke & Hollenbach's theoretical simulations (1983, Fig. 5). The trapping fraction of gas particles upon solid surfaces is defined by Burke & Hollenbach (1983) as the ratio of the flux of gas particles trapped on the surface by the flux of gas particles of number density n. One can see that their results do not match our estimations and other predictions (over the complete range of gas temperatures). This is certainly due to their model for silicate surfaces.

The formulae $S = (k_B T/E_0 + 1)^{-2}$ used by Buch & Zhang (1991) for the sticking of H atoms on amorphous water ice can match our results on silicate surfaces if we substitute the energetic parameter $E_0/k_B = 102$ K obtained by Buch & Zhang (1991) (for water ice) by the new parameter $E_0/k_B = 80$ K. Nevertheless, none of these simplified formulae deal with the isotopic effect between D and H atoms. More importantly, the ratio between S(D) and S(H) is higher than 1.5 for gas temperatures $T > 50$ K and increases up to 2 at about 300 K (see Figure 3). Furthermore, the sticking curve S(HD) estimated from our formulae (1) with the parameters of Table 1 exhibits the following properties: (a) for $T > 40$ K, S(HD) is slightly higher than S(D), (b) the ratio S(HD)/S(D) starts to be greater than 1.5 for $T > 200$ K.

Even if the absolute value of $T_0(H) = (25 \pm 10)$ K can be discussed, the isotopic effect based on our measurements and on independent calculations has been demonstrated. When taking $T_0(H) = (35 \pm 10)$ K (corresponding to our upper value when $T_0(D_2)/T_0(D) = 1.67$) instead of $T_0(H) = (25 \pm 10)$ K (value used here), the sticking coefficient values of H atoms increased slightly except for zero gas temperature. Consequently, the isotopic effect between H and D atoms prevails because the ratio between S(D) and S(H) is higher than 1.2 for gas temperatures $T > 50$ K and reaches $\sim 1.6$ at around 300 K. In contrast, if we consider for example that $S_0(D) = 0.8$ instead of unity and that the accommodation of D atoms is less efficient on silicates at zero temperature, the sticking coefficient value of D atoms decreases notably for gas temperatures $T < 300$ K, making the isotopic effect between D and H atoms less significant than previously noted.

## 5. Conclusions

We have experimentally studied the molecular gas temperature dependence of the sticking coefficient of $H_2$ and $D_2$ molecules on silicate grain surfaces held at 10 K. These results were well fitted with the model developed in our previous paper, giving the sticking coefficient $S(T)$ of molecules on amorphous water ice. Using formulae (1), we were able to provide the sticking parameters $S_0$ and $T_0$ for $D_2$ and $H_2$. Using an experimental benchmark, we proposed values for D, H, and HD species on silicate interstellar grains. Finally, we proposed a complete set of parameters both for amorphous solid water ices and for amorphous silicates, which can be implemented easily in astrochemical models. We suggest that the difference in the measured sticking coefficients has important consequences on the estimations of isotopic fractionation.

Acknowledgements. The authors acknowledge the financial support from the ANR (Agence Nationale de la Recherche, contract 07-BLAN-0129), the Conseil Régional d'Ile de France (SESAME contract I- 07-597R), and the Conseile Général du Val d'Oise as well as from the European Community FP7- ITN-e e Marie-Curie Programme (LASSIE project, grant agreement #238258), and the French National PCMI program funded by the CNRS.